\documentclass[prd,twocolumn,floatfix,preprintnumbers,showpacs]{revtex4}
\usepackage{graphicx}
\usepackage{dcolumn}
\usepackage{bm}
\usepackage{color}

\def\lsim{\mathrel{\mathop
  {\hbox{\lower0.5ex\hbox{$\sim$}\kern-0.8em\lower-0.7ex\hbox{$<$}}}}}
\def\gsim{\mathrel{\mathop
  {\hbox{\lower0.5ex\hbox{$\sim$}\kern-0.8em\lower-0.7ex\hbox{$>$}}}}}

\begin{document}
\newcommand{\mincir}{\raise
-2.truept\hbox{\rlap{\hbox{$\sim$}}\raise5.truept 
\hbox{$<$}\ }}
\newcommand{\magcir}{\raise
-2.truept\hbox{\rlap{\hbox{$\sim$}}\raise5.truept
\hbox{$>$}\ }}
\newcommand{\minmag}{\raise-2.truept\hbox{\rlap{\hbox{$<$}}\raise
6.truept\hbox
{$>$}\ }}

\newcommand{\half}{{1\over2}}
\newcommand{\bk}{{\bf k}}
\newcommand{\Ocdm}{\Omega_{\rm cdm}}
\newcommand{\ocdm}{\omega_{\rm cdm}}
\newcommand{\OM}{\Omega_{\rm M}}
\newcommand{\OB}{\Omega_{\rm B}}
\newcommand{\oB}{\omega_{\rm B}}
\newcommand{\OX}{\Omega_{\rm X}}
\newcommand{\cltt}{C_l^{\rm TT}}
\newcommand{\clte}{C_l^{\rm TE}}
\newcommand{\mwdm}{m_{\rm WDM}}
\newcommand{\mnu}{\sum m_{\rm \nu}}
\newcommand{\etal}{{\it et al.~}}
\newcommand{\lya}{{Lyman-$\alpha$~}}
\input epsf

\preprint{DFPD 05/A/08, LAPTH-1086/05, hep-ph/0501562}
\title{
Constraining  Warm Dark Matter candidates
including sterile neutrinos \\
and light gravitinos 
with WMAP and the Lyman-$\alpha$ forest}
\author{Matteo Viel,$^1$ 
Julien Lesgourgues,$^{2,3}$
Martin G.~Haehnelt,$^1$ 
Sabino Matarrese,$^{4,3}$ 
Antonio Riotto$^3$}
\affiliation{
$^1$Institute of Astronomy, Madingley Road, Cambridge CB3 0HA\\
$^2$Laboratoire de Physique Th\'eorique LAPTH, F-74941
Annecy-le-Vieux Cedex, France\\
$^3$ INFN, Sezione di Padova,
Via Marzolo 8, I-35131 Padova, Italy\\
$^4$ Dipartimento di Fisica ``G. Galilei'', Universit\`a di Padova,
Via Marzolo 8, I-35131 Padova, Italy
}
\date{\today}
\pacs{98.80.Cq}
\begin{abstract}
The matter power spectrum at comoving scales of $(1-40) \, h^{-1}
\mathrm{Mpc}$ is very sensitive to the presence of Warm Dark Matter
(WDM) particles with large free streaming lengths.
We present constraints on the mass of WDM particles
from a  combined analysis of the matter power spectrum inferred from the large
samples of high resolution high signal-to-noise Lyman-$\alpha$ forest
data of Kim \etal (2004) and Croft \etal (2002) and  the
cosmic microwave background data of WMAP. We obtain a lower limit of $\mwdm
\gsim 550$~eV ($2\sigma$) for early decoupled thermal relics and
$\mwdm \gsim 2.0$ ~keV ($2\sigma$) for sterile neutrinos.  We also
investigate the case where in addition to cold dark matter a light
thermal gravitino 
with fixed effective temperature
contributes significantly to the matter density. In
that case the gravitino density is proportional to its mass, and we
find an upper limit $m_{3/2}\lsim 16$~eV ($2\sigma$). This
translates into a bound on the scale of supersymmetry breaking,
$\Lambda_{\rm susy}\lsim 260$ TeV,  for models of 
supersymmetric gauge mediation in which the gravitino is 
the lightest supersymmetric particle.

\end{abstract}

\maketitle
 
\section{Introduction}

It is now observationally well established that the Universe is close
to flat and that matter accounts for about (25-30) percent of the
total energy density most of which must be in the form of dark matter (DM)
particles. Candidates of dark matter particles are generally
classified according to their velocity dispersion which defines a
free-streaming length. On scales smaller than the free-streaming
length, fluctuations in the dark matter density are erased and
gravitational clustering is suppressed.  The velocity dispersion of Cold
Dark Matter (CDM) particles is by definition so small that the
corresponding free-streaming length is irrelevant for cosmological
structure formation. That of Hot Dark Matter (HDM), e.g. light neutrinos, is
only one or two orders of magnitude smaller than the speed of light,
and smoothes out fluctuations in the total matter density even on
galaxy cluster scales, which leads to strong bounds on their mass and
density \cite{wmap,neutrinos,Hannestad:2003ye}. 
Between these two limits, there exists an
intermediate range of dark matter  candidates generically called Warm Dark Matter
(WDM). If such particles are  initially in thermal equilibrium, they
decouple well before ordinary neutrinos. As a result their temperature is  
smaller and their free-streaming length shorter than that of ordinary
neutrinos. For instance, thermal relics with a mass of order
$1$~keV and a density  $\Omega_x\sim 0.25$
would have a free-streaming length comparable to galaxy 
scales ($\lambda_\mathrm{FS}\sim 0.3$~Mpc). 

There exist many WDM candidates whose origin is firmly rooted
in particle physics. A prominent example  is the gravitino, the supersymmetric partner of the
graviton. The gravitino  mass $m_{3/2}$ is 
generically of the order of $\Lambda_{\rm susy}/M_p$, where 
$\Lambda_{\rm susy}$ is the
scale of supersymmetry breaking and $M_p$ is the Planckian scale. 
If $\Lambda_{\rm susy}$ is large the Lightest Supersymmetric Particle (LSP),
which is then not the gravitino, decouples in the non-relativistic regime
and provides a viable CDM candidate.
If, however,  $\Lambda_{\rm susy}\lsim 
10^6$ GeV, as predicted by theories where supersymmetry breaking is
mediated by gauge interactions, the 
gravitino is likely to be the LSP. Such a light gravitino 
has a wide range of possible masses (from $10^{-6}$eV up to the keV region).
The gravitino -- or better to say its helicity 1/2 component --
decouples when it is still relativistic. At this time  the 
number of degrees of freedom in the Universe is typically of order
one hundred, much larger than at neutrino decoupling. The effective gravitino
temperature is therefore always  smaller than the neutrino temperature,
and such light gravitinos can play the role of WDM \cite{gravitinoWDM}.
Their velocity dispersion is non-negligible during the time of
structure formation, but  smaller than the velocity dispersion of
active neutrinos with the same mass.
 
Other particles may decouple even earlier from thermal equilibrium
while still relativistic and may act as warm
dark matter.  The same is true for right-handed or sterile neutrinos
added to the standard electroweak theory. Since
their only direct coupling is to left-handed or active neutrinos, the
most efficient production mechanism is via neutrino oscillations. If
the production rate is always less than the expansion rate, then these
neutrinos will never be in thermal equilibrium.  However, enough of
them may be produced to account for the observed matter density. 
Right-handed (sterile) neutrinos  with a mass  of order keV are therefore natural 
WDM candidates \cite{sterileWDM,CDW}.

The suppression of structures on scales of a Mpc and smaller, 
in WDM models with a particle mass of $\sim$ 1 keV,  has been invoked 
as a possible  solution to two major apparent  conflicts between CDM models 
and  observations in the local Universe (see \cite{bode,Moore:1999gc,avila,nara} and 
references therein). The first problem concerns the inner mass density 
profiles of simulated dark matter halos that are generally 
more cuspy than inferred from the rotation curves of many dwarfs 
and low surface brightness galaxies. Secondly,  N-body simulations of 
CDM models predict large numbers of low mass halos greatly in excess
of the observed number of satellite galaxies in the Local Group. 
It appears, 
however, difficult to find WDM parameters which solve all apparent CDM problems 
simultaneously and it is also not  clear whether the tension between 
theory and observations is due to an inappropriate comparison 
of numerical simulations and observational data. The main difficulty
is that the modeling of the inner  density profile of galaxies and 
the abundance of satellite galaxies requires simulations on 
scales where the matter density field is highly non-linear at low
redshift. Furthermore,  many of the relevant baryonic processes are 
poorly understood. 

At $z\sim (2-3)$ the gravitational clustering of the matter
distribution spectrum at scales larger than the free streaming length
of a thermal $1\,\mathrm{keV}$ WDM particle is still in the mildly
non-linear regime. At these redshift high-resolution \lya forest
data offer an excellent probe of the matter power spectrum at these
scales. Narayanan et al. \cite{nara} compared the flux power spectrum
and flux probability distribution of a small sample of high resolution
quasar (QSO) absorption spectra to that obtained from numerical dark matter
simulations.  In this way they were able to constrain the mass of a
thermal WDM particle (assumed to account for all the dark matter in
the Universe) to be $m_\mathrm{WDM} \gsim 750 \, \mathrm{eV}$.

We will use here the linear DM power spectrum inferred from two large
samples of QSO absorption spectra \cite{kim04,croft} using
state--of--the--art hydrodynamical simulations \cite{vhs} combined
with cosmic microwave background data from WMAP in order to obtain new
constraints on the mass of possible thermally distributed WDM
particles. We thereby assume that WDM particles account for all the DM in our
Universe and we do not put any priors on their temperature. We will also further
investigate constraints on the mass of light gravitinos with
realistic effective temperature.

The paper is organized as follows. In Section II we will describe 
the effect of WDM on the power spectrum of the matter 
distribution. In Section III we will summarize the method used in inferring 
the linear matter power spectrum  from  Lyman-$\alpha$ forest
observations  and discuss the systematic errors involved. 
In Section IVA we will derive a lower limit for the mass of the dark 
matter particle in a pure $\Lambda$WDM
model.  In Section IVB, we will obtain an upper limit for 
the mass of light gravitinos  in a mixed $\Lambda$CWDM universe.  Section V
contains our conclusions and some comments on the implications of our
results for particle physics.

\section{WDM signatures on the matter power spectrum}

Standard neutrinos are believed to decouple from the plasma when they
are still relativistic, slightly before electron-positron
annihilation.  They will keep the equilibrium distribution of a
massless fermion with temperature $T_{\nu} \simeq (4/11)^{1/3}
T_{\gamma}$. Their total mass summed over all neutrino species 
and their density are related through
the well-known relation $\Omega_{\nu}h^2 \simeq (m_{\nu} / 94
\mathrm{eV})$ in the instantaneous decoupling limit. 
This result does, however,  not apply to WDM for which: (i) either
decoupling takes place earlier,  (ii) or the particles were
never in thermal equilibrium. 

If decoupling takes place 
at temperature $T_D$ while there are  $g_*(T_D)$ degrees of freedom
the particles do not share the entropy release from the 
successive annihilations and their temperature is 
suppressed today by a factor
\begin{equation}
{T_{x} \over T_{\nu}} = \left( {10.75 \over g_*(T_D)} \right)^{1/3} < 1~,
\label{tempratio}
\end{equation}
as is the case e.g. for gravitinos.  If the particles have never been
in thermal equilibrium, there is a wide range of  possible models. We will
restrict our analysis here  to the case of sterile neutrinos created from
oscillations with active neutrinos, which share the same phase--space
distribution but with an overall suppression factor $\chi$. In both
cases, the number density of WDM is smaller than that of standard
neutrinos, and the same present-day non-relativistic energy density
corresponds to a larger particle mass,
\begin{equation}
\omega_x = \Omega_{x} h^2 = \beta \left({m_{x} \over 94 \, \mathrm{eV}}\right),
\label{omegawdm}
\end{equation}
with $\beta=(T_{x} / T_{\nu})^{3}$ for early decoupled thermal relics and
$\beta=\chi$ for sterile neutrinos.
Due to their free-streaming velocity, WDM particles  slow down the 
growth of structure and  suppress the total matter power spectrum $P(k)$
on scales smaller than their free-streaming scale which is roughly
given by
\begin{equation}
k_{\rm FS} = {2\pi \over \lambda_{\rm FS}} \sim
5 \, \mathrm{Mpc}^{-1} 
\left({m_{x} \over 1 \, \mathrm{keV}}\right)
\left({T_{\nu} \over T_{x}}\right)~,
\end{equation}  
see e.g.~\cite{Bond:1980ha}. Note, however, that there exist different
definitions of the free-streaming or damping scale.  The effect
of the free streaming on the matter distribution can be described by a
relative ``transfer function'',
\begin{equation}
T(k)=[P(k)_{\Lambda\mathrm{WDM}}/P(k)_{\Lambda\mathrm{CDM}}]^{1/2}
\label{transdef}
\end{equation}
which is simply the square root of the ratio of the matter power
spectrum in the presence of WDM to that in the presence of purely cold dark
matter, for fixed  cosmological parameters. 

Since we allowed for an arbitrary normalization
factor $\chi$, the three parameters $\omega_{x}$, $m_{x}$ and $T_{x}$
are independent.  Note, however,  that the evolution
equations for WDM as well as the corresponding terms in the Einstein
equations can be reparametrized entirely in terms of two parameters,
for instance $\omega_{x}$ and $m_{x}/T_{x}$ (where $T_x$ is expressed
in units of the neutrino or photon temperature).  There is a simple 
one--to--one correspondence between the masses of thermal WDM particles
and sterile neutrinos 
for which the effect on the matter distribution 
and thus the transfer function  for both models are identical~\cite{CDW}.  
Writing equation~(\ref{omegawdm}) in the two cases and equating 
$\omega_{x}$ and  $m_{x}/T_{x}$, one obtains
\begin{equation}
m_\mathrm{sterile \, \nu}
=  4.43 \, \mathrm{keV}
\left({m_\mathrm{thermal} \over 1 \, \mathrm{keV}}\right)^{4/3}
\!
\left({0.25 (0.7)^2 \over \omega_{x}}\right)^{1/3}.
\label{onetoone} 
\end{equation}
In the following, we will thus consider only the thermal
model and simply translate any mass limit into that for the 
sterile neutrinos using eq.~(\ref{onetoone}).

Let us now discuss briefly some simple scenarios. 

\vspace{0.3cm}

{\it Pure warm dark matter models}.
In the case in which the Universe contains only WDM (in
addition to the usual baryon, radiation and cosmological constant components),
the transfer function can be approximated by the fitting 
function~\cite{bode}
\begin{equation}
T(k)=[1+(\alpha k)^{2 \nu}]^{-5/\nu}
\label{Tshape}
\end{equation}
where $\alpha$, the scale of the break, is a function of the WDM parameters,
while the index $\nu$ is fixed.
With a Boltzmann code simulation
(using either {\sc cmbfast}~\cite{Seljak:1996is} or 
{\sc camb}~\cite{Lewis:1999bs}), 
we find that $\nu=1.12$ is the best fit for $k<5 \, h$Mpc$^{-1}$
and we obtain
\begin{eqnarray}
\alpha \! &=& \!
0.24
\left({m_{x}/T_{x} \over 1 \, \mathrm{keV} / T_{\nu}} \right)^{\!-0.83}
\!\!
\left({\omega_{x} \over 0.25 (0.7)^2}\right)^{\!-0.16} 
\!\!\!\!\!
\mathrm{Mpc}
\nonumber \\
&=& \! 
0.049
\left({m_{x} \over 1 \, \mathrm{keV}}\right)^{\!-1.11}
\!\!
\left({\Omega_{x} \over 0.25}\right)^{\!0.11}
\!\!
\left({h \over 0.7}\right)^{\!1.22}
\!\!\!\!\!\!\!
h^{-1}\mathrm{Mpc}
\label{fit}
\end{eqnarray}
where the second line applies only to the case of thermal relics.
This expression is close to that  of~\cite{bode} 
(and exactly identical
to the results of \cite{Hansen:2001zv}, 
except for the front factor which was
misprinted in this reference).

\begin{figure}[h!]
\begin{center}
\hspace{-0.9cm}
\includegraphics[angle=0,width=5.22cm]{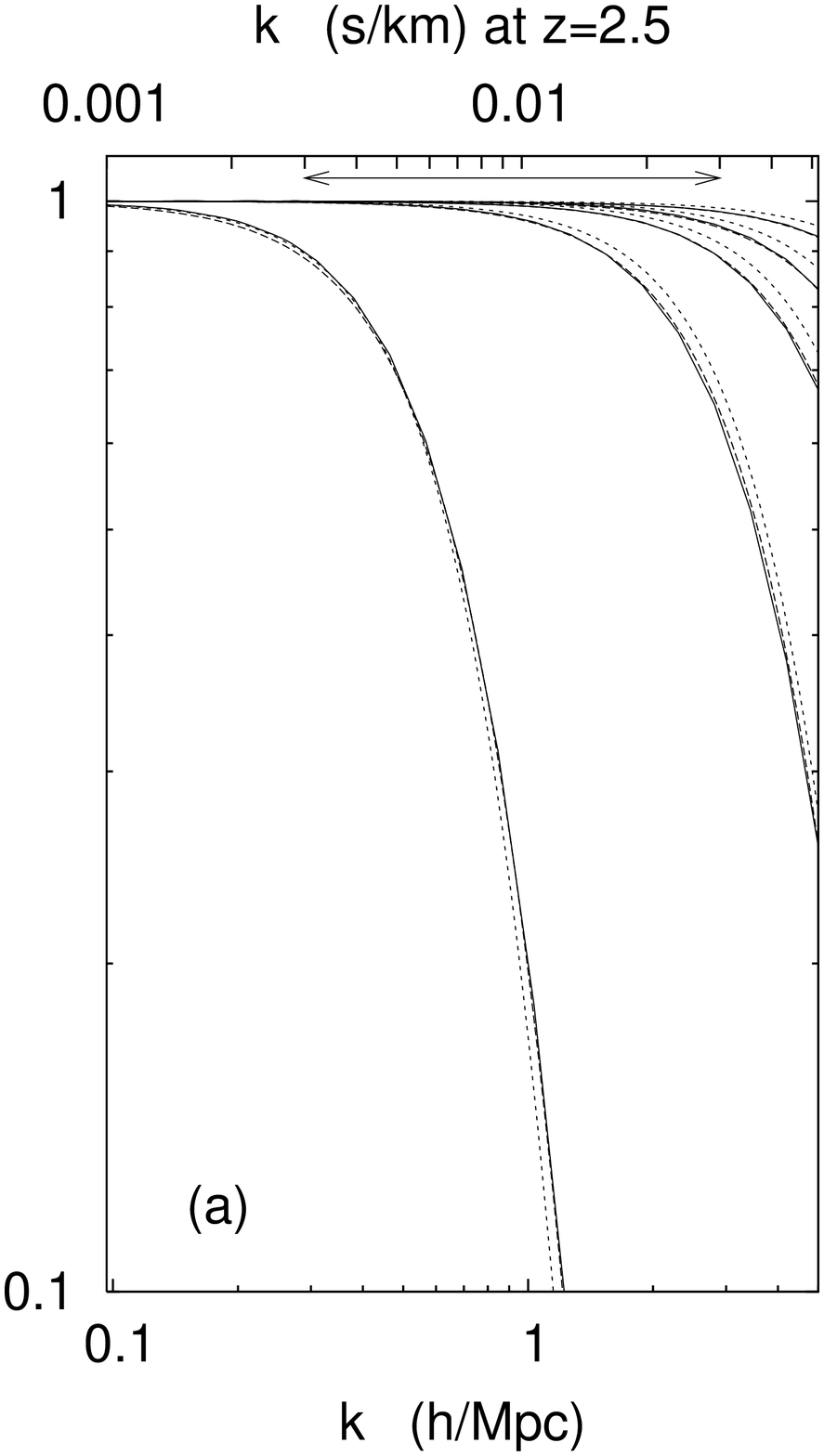}
\hspace{-1.2cm}
\includegraphics[angle=0,width=5.22cm]{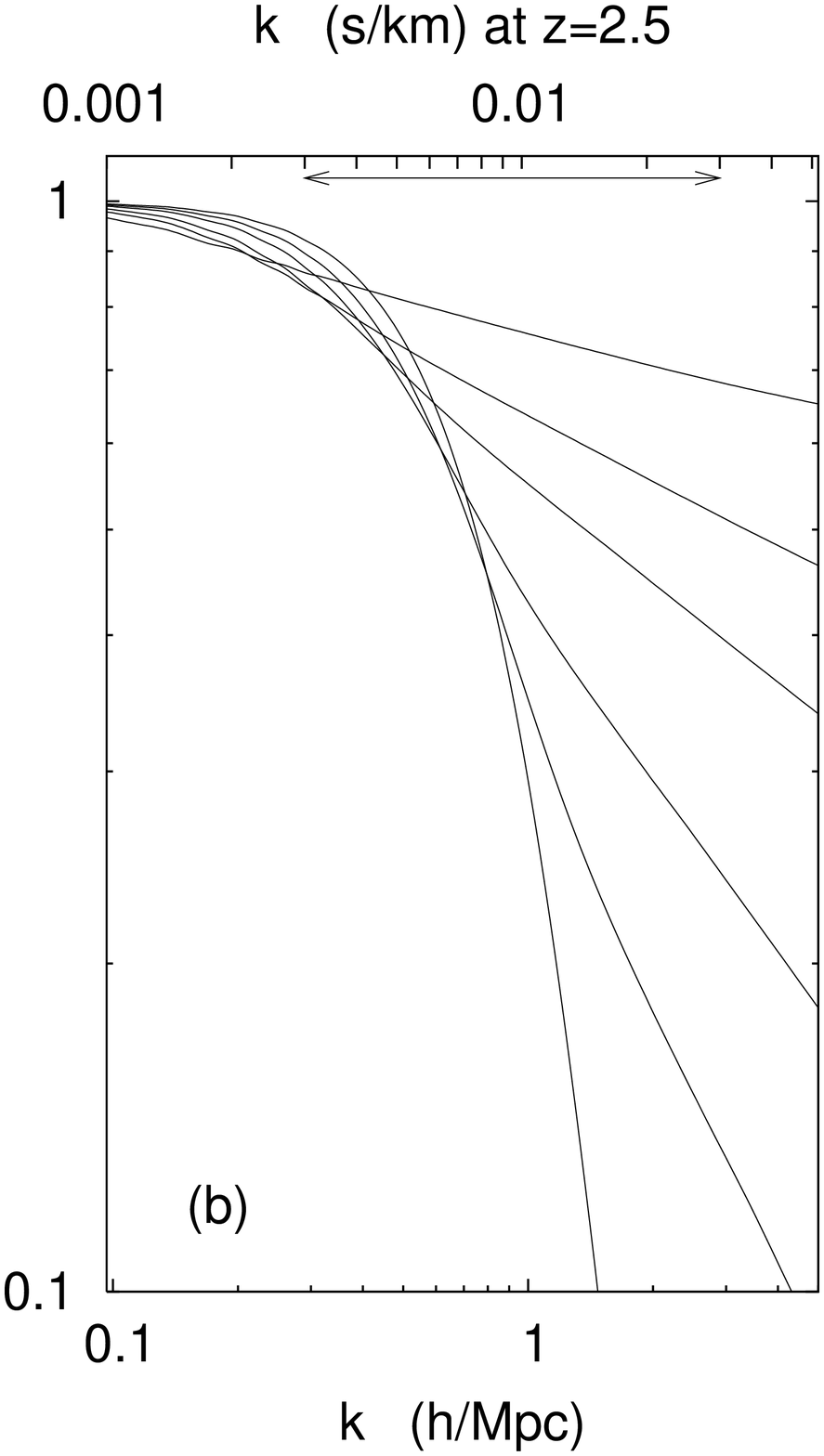}
\end{center}
\caption{\label{fig1} 
The WDM transfer function $T(k)$ defined in equation~(\ref{transdef})
for various models (all  with the same cosmological parameters
$\Omega_\mathrm{B}\!=\!0.05$, $\Omega_\mathrm{DM}\!=\!0.25$, 
$\Omega_{\Lambda}\!=\!0.70$).
(a) Pure $\Lambda$WDM model with, from left to right, 
$T_x/T_{\nu}=0.5,0.3,0.25,0.226,0.2$, corresponding to
$m_x = 92, 427, 738, 1000, 1441 \, \mathrm{eV}$. The solid curves
are obtained numerically, the long--dashed curves
(essentially indistinguishable from the solid curves) show the analytical
fits based on equation~(\ref{fit}),  the short--dashed curves
those based on Ref.~\cite{bode}.
(b) Mixed $\Lambda$CWDM model 
for a warm component which decoupled when
$g_*(T_D)=100$, like {\it e.g.} a light  gravitino (with mass  proportional to
density and $\Omega_\mathrm{CDM}=0.25-\Omega_x$). The solid curves
show the numerical results for $m_x=10,20,30,50,70,100\, \mathrm{eV}$, from
top right to bottom left. At  the top-axis we show the wavenumber scale
in s/km (assuming the above cosmological model and  $z=2.5$
intermediate between the two \lya data sets analysed here). In both
panels the double arrow indicates the range of wavenumbers used in our analysis.}
\end {figure}

In Figure \ref{fig1}, we plot the results of the numerical simulations and compare them 
with our  analytical fit. The power spectrum of the $\Lambda$WDM  
model differs from that of the corresponding $\Lambda$CDM model only at small scales. 
This is the reason why WDM has been suggested as a solution to the apparent 
``crisis'' of CDM models on small scales. The matter density is fixed by the CMB and other
data  to $\omega_x \sim 0.12$. For pure $\Lambda$WDM  models
for which $m_x \sim 1 \, \mathrm{keV}$ this gives $g_*(T_D) \sim 10^3$. 
The particles therefore decouple extremely early in such a model. 
Note that the light gravitino suggested by gauge mediated 
models of supersymmetry breaking  decoupled when $g_*$ was much smaller, 
of the order of one hundred.  

\vspace{0.3cm}

{\it Mixed models with CDM and gravitinos (or any thermal relics with
$g_*(T_D)$ of a hundred)}. We will now consider the case of thermal
relics which decoupled at a temperature when $g_*(T_D) \sim 100$, 
as {\it e.g.} the light gravitino, which is likely to be the 
LSP in gauge-mediated
theories of supersymmetry breaking. Note that $g_*(T_D) \sim 100$
corresponds to a  decoupling temperature  between a few GeV and a few
TeV (possibly much higher, depending on which theory is assumed beyond the standard model). 
Note further that the temperature of
thermal relics with   $g_*(T_D) \sim 100$ is about a factor two smaller 
than that of ordinary neutrinos. They are thus not equivalent to 
HDM even in the small mass limit. 
For fixed $g_*$, both the
density $\omega_x$ and the free-streaming wavenumber are proportional
to the mass $m_x$. Taking $g_*=100$, one gets
\begin{equation}
\omega_x = 0.117 \, {m_x \over 100 \, \mathrm{eV}}~, \qquad
k_{\rm FS} \sim 1.5 \, h \mathrm{Mpc}^{-1} {m_x \over 100 \, \mathrm{eV}}~.
\label{gravitino}
\end{equation}
In the absence of any cold dark matter, the preferred value $\omega_x
\sim 0.12$ would lead to a mass of order 100~eV and to a complete
smoothing out of fluctuations for $k > 1.5 \, h \mathrm{Mpc}^{-1}$, at
odds with the Lyman-$\alpha$ forest data (and with various other constrains
related to structure formation~\cite{Pierpaoli:1997im}).  However, in
the presence of CDM, small--scale structures are not completely
suppressed and the free-streaming effect leads to a step--like transfer
function (as for massive neutrinos, but with a step at a much smaller
scale), while in the small $m_x$ limit standard $\Lambda$CDM is
recovered. For a light gravitino in the the mass range $10
\,\mathrm{eV} \lsim m_{3/2} \lsim 100 \,\mathrm{eV}$,
$g_*(T_D)$ varies
between 87 and 101 depending on the details of the supersymmetric
model~\cite{Pierpaoli:1997im}.  As we already pointed out a light
gravitino with larger mass would result in a matter density which
exceeds that observed and as we will show later smaller masses are
hard to constrain with current observations.  In Figure 1 we show the
transfer function computed numerically for a few values of the mass in
the range $(10 - 100)$~eV, corresponding to $k_{\rm FS}$ in the range
$(0.15 - 1.5)\, h\mathrm{Mpc}^{-1}$.  Since the matter power spectrum
measured from galaxy redshift surveys can be compared to linear
predictions up to $k \sim 0.15 \, h \mathrm{Mpc}$, it has little
sensitivity to this model but as we will show in Section IV the matter
power spectrum inferred from Lyman-$\alpha$ data provides a tight upper
limit on the gravitino mass.

\vspace{0.3cm}

{\it Mixed models with warmer thermal relics}.
Finally, the WDM particles could decouple at temperature of order
1 GeV or below (i.e. just before or after the QCD phase transition),
when the number of degrees of freedom was in the range
$10.75 < g_* < 80$ leading later to a temperature
$0.5 < (T_x/T_{\nu}) < 1$.
This case smoothly interpolates between that of
ordinary neutrinos (HDM) and that of the previous paragraph.  The
free-streaming scale is smaller than for the gravitino and
these models have  been  constrained by the matter power spectrum
reconstructed from galaxy redshift surveys \cite{Hannestad:2003ye}.  
We will not consider this case here.

\vspace{0.3cm}

\section{Probing the matter power spectrum with the Lyman-$\alpha$
forest in QSO absorption spectra}
 
It is well established by analytical calculation and hydrodynamical 
simulations that the \lya forest blueward of the \lya emission line 
in QSO spectra  is produced by the inhomogenous distribution of a
warm  ($\sim 10^4$ K) and photoionized intergalactic medium (IGM)
along the line of sight.  The opacity  fluctuations in the spectra 
arise from fluctuations in the matter density and trace the
gravitational clustering of the matter distribution in the
quasi-linear regime \cite{bi}. The \lya forest has thus been used 
extensively  as a probe of the matter power spectrum on comoving scales 
of  $(1-40) \, h^{-1}$Mpc \cite{bi,croft,vhs,mcdonald2}.

The \lya optical depth in  velocity space $u$ (km/s) is 
related to the neutral hydrogen distribution in real space 
as (e.g. 
Ref. \cite{hgz}):
\begin{equation}
\tau(u)={\sigma_{0,\alpha} ~c\over H(z)} \int_{-\infty}^{\infty}
dy\, n_{\rm HI}(y) ~{\cal V}\left[u-y-v_{\parallel}(y),\,b(y)\right]dy \;,
\label{eqtau2} 
\end{equation} 
where $\sigma_{0,\alpha} = 4.45 \times
10^{-18}$ cm$^2$ is the hydrogen Ly$\alpha$ cross-section, $y$ is the real-space coordinate
(in km s$^{-1}$), ${\cal V}$ is the standard Voigt profile normalized
in real-space, $b=(2k_BT/mc^2)^{1/2}$ is the velocity dispersion in
units of $c$, $H(z)$ the Hubble parameter,
$n_{\rm HI}$ is the local density of neutral hydrogen and
$v_{\parallel}$ is the peculiar velocity along the line-of-sight. The
density of neutral hydrogen can be obtained by solving the
photoionization equilibrium equation (e.g. \cite{katz}). 
The neutral hydrogen in the IGM responsible for the \lya forest
absorptions is highly ionized due to the metagalactic
ultraviolet (UV) background radiation produced by stars and QSOs at
high redshift. This optically thin gas in photoionization equilibrium
produces a \lya optical depth of order unity.

The balance between the photoionization heating by the UV background and adiabatic 
cooling by the expansion of the universe drives most of the
gas with $\delta_b< 10$, which dominates the \lya opacity, 
onto a power-law density relation $T=T_0\,(1+\delta_b)^{\gamma-1}$, where
the parameters $T_0$ and $\gamma$
depend on the reionization history and spectral shape of the UV
background and $\delta_b$ is the local gas overdensity ($1+\delta_b=\rho_b / \bar{\rho}_b$).

The relevant physical processes can be readily modelled in hydro-dynamical 
simulations.  The physics of a photoionized IGM that traces the dark
matter distribution is, however, sufficiently simple that 
considerable insight can be gained from analytical 
modeling of the IGM opacity  based on the so called Fluctuating 
Gunn Peterson Approximation neglecting the effect of peculiar
velocities and the thermal broadening \cite{fgpa}. 
The Fluctuating  Gunn Peterson Approximation makes use of the
power-law temperature density relation and describes 
the relation between \lya opacity and gas density (see \cite{rauch,croft})
along a given line of sight as follows,
\begin{eqnarray}
\tau(z) &\propto & (1+\delta_b(z))^2 \, T^{-0.7}(z) 
~=~ {\cal A}(z) \, (1+\delta_{b}(z))^\beta~, 
\label{eqtau} \\
{\cal A}(z) & = & 0.433
\left(\frac{1+z}{3.5}\right)^6 
\left(\frac{\Omega_{b} h^2}{0.02}\right)^2
\left(\frac{T_0}{6000\;{\rm K}}\right)^{-0.7} \;\times \nonumber \\
& & \left(\frac{h}{0.65}\right)^{-1}
\left(\frac{H(z)/H_0}{3.68}\right)^{-1} 
\left(\frac{\Gamma_{\rm HI}}
{1.5\times 10^{-12}\;{\rm s}^{-1}}\right)^{-1}\; ,\nonumber
\end{eqnarray}
where $\beta \equiv 2 - 0.7\, (\gamma-1)$ in the range $1.6-1.8$,
$\Gamma_{\rm HI}$  the HI photoionization rate, $H_0=h\, 100$ km/s/Mpc the Hubble
parameter at redshift zero. 
For a quantitative anlysis, however, full hydro-dynamical simulations,  
which  properly simulate  the non-linear evolution of the
IGM and its thermal state, are needed.

Equations (\ref{eqtau2},\ref{eqtau}) show how the  observed flux
$F=\exp{(-\tau)}$ depends on the underlying local gas density $\rho_b$, 
which in turn is simply related to the dark matter density, at least 
at large scales where the baryonic pressure can be neglected \cite{gh}. Statistical
properties of the flux distribution, such as the flux power spectrum,
are thus closely related to the statistical properties of the underlying
matter density field.

\subsection{The data: from the quasar spectra to the flux power spectrum}

The power spectrum of the observed flux in high-resolution \lya forest
data provides meaningful constraints on the dark matter power spectrum
on scales of $0.003\,{\rm s/km} < k < 0.03\,{\rm s/km}$, roughly
corresponding to scales of $(1-40) \, h^{-1}$Mpc (somewhat dependent on
the cosmological model). At larger scales the errors due to
uncertainties in fitting a continuum (i.e. in removing the long
wavelength dependence of the spectrum emitted by each QSO) become
very large while at smaller scales the contribution of metal
absorption systems becomes dominant (see e.g. \cite{kim04,mcdonald}).
In this paper, we will use the dark matter power spectrum that Viel,
Haehnelt \& Springel~\cite{vhs} (VHS) inferred from the flux power
spectra of the Croft et al.~\cite{croft} (C02) sample and the LUQAS
sample of high-resolution \lya forest data \cite{luqas}.  The C02
sample consists of 30 Keck high resolution HIRES spectra and 23 Keck
low resolution LRIS spectra and has a median redshift of $z=2.72$.
The LUQAS sample contains 27 spectra taken with the UVES spectrograph
and has a median redshift of $z=2.125$.  The resolution of the spectra
is 6 km/s, 8 km/s and 130 km/s for the UVES, HIRES and LRIS spectra,
respectively.  The S/N per resolution element is typically
30-50. Damped and sub-damped \lya systems have been removed from the
LUQAS sample and their impact on the flux power spectrum has been
quantified by \cite{croft}.  Estimates for the errors introduced by
continuum fitting, the presence of metal lines in the forest region
and strong absorptions systems have also been 
made~\cite{mcdonald,croft,hui,kim04}.

We stress that the total sample that went into the estimate of the
linear power spectrum used here is about 10 times larger than
the one used in the study by \cite{nara}.

\subsection{From the flux power spectrum to the linear
matter power spectrum}
\label{hydro}

\begin{figure*}[]
\vspace{1cm}
\includegraphics[angle=0,width=18.0cm,height=7.0cm]{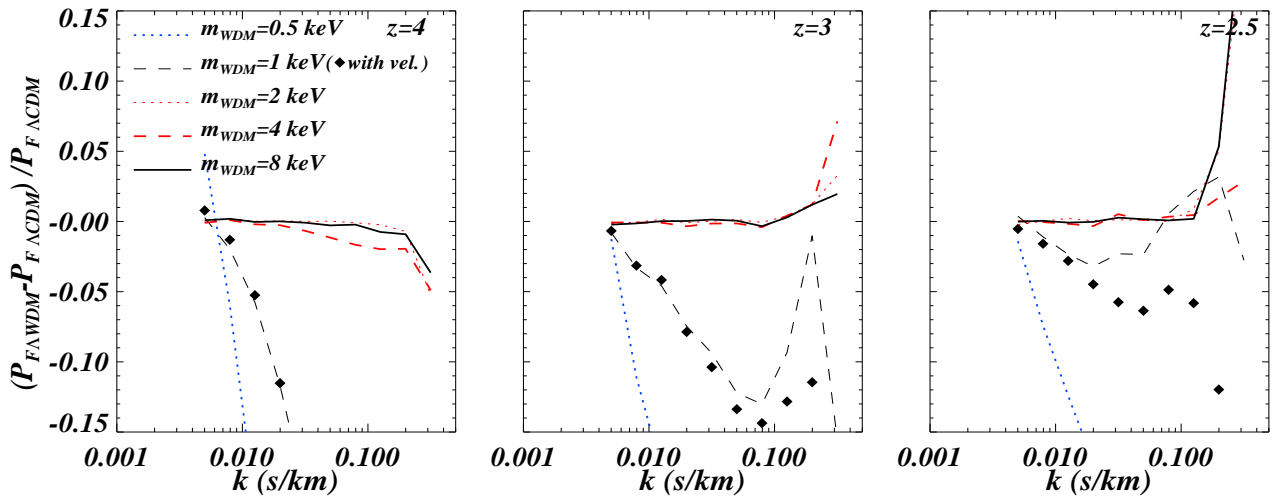}
\caption{\label{fig2} Fractional difference of the flux power spectrum
for a hydro-dynamical simulation of a $\Lambda$WDM model and the flux power
spectrum for a $\Lambda$CDM model.  The  simulations have a box size of 30
$h^{-1}$\,Mpc and 200$^3$ gas and 200$^3$ dark matter particles. The
results are for three different redshifts ($z=4,3,2.5$, respectively
from left to right) and for five different values of the (thermal WDM)
particle mass: 0.5 keV (thick dotted), 1 keV (thin dashed), 2 keV
(thin dotted), 4 keV (thick dashed) and 8 keV (continuous). The
diamonds show the results for a simulation with a WDM particle of mass
1keV when random oriented velocity dispersion drawn from a Fermi-Dirac
distribution have been added in the initial conditions.}
\end{figure*}

VHS have used numerical simulation to calibrate the relation between 
flux power spectrum and linear dark matter power spectrum  with
a method proposed by C02 and improved by \cite{gnedham}
and VHS.  A set  of hydro-dynamical simulations for a coarse
grid of the relevant parameters is used  to find a model that  provides
a reasonable  but not exact fit to the observed flux power
spectrum. It is then  assumed that the differences
between the model and the observed linear power spectrum  depend linearly 
on the matter power spectrum.  

The hydro-dynamical simulations are used to define a bias function
between flux and matter power spectrum: $ P_F(k) = b^2(k)\;P(k)$, on
the range of scales of interest. In this way the linear matter power spectrum
can be recovered with reasonable computational resources.  This method
has been found to be robust provided the systematic uncertainties 
are properly taken into account \cite{vhs,gnedham}.
Running hydro-dynamical simulations for a fine grid of all the relevant
parameters is unfortunately computationally prohibitive.

The  use of state-of-the-art hydro-dynamical simulations is a significant 
improvement  compared to previous studies which used 
numerical simulation of dark matter only \cite{croft,nara}.
We use a new version of the parallel TreeSPH code GADGET
\cite{volker} in its TreePM mode which speeds up the
calculation of long-range gravitational forces considerably. The
simulations are performed with periodic boundary conditions with an
equal number of dark matter and gas particles. Radiative cooling and
heating processes are followed using an implementation similar to
\cite{katz} for a primordial mix of hydrogen and helium. The UV
background is given by \cite{haardt}. In order to maximise the speed of
the simulation a simplified criterion of star formation has been
applied: all the gas at overdensities larger than 1000 times the mean
overdensity is turned into stars \cite{vhs}.
The simulations were run on {\sc cosmos}, a 152 Gb shared memory Altix 3700
with 152 CPUs hosted at the Department of Applied Mathematics and
Theoretical Physics (Cambridge).

In order  to check the impact of a WDM particle on the flux
power spectrum, we have run a set of simulations with different WDM
particle masses and compared them with a $\Lambda$CDM simulation with the
same phases of the initial conditions. The cosmological parameters for both simulations are 
$\Omega_{{\rm M}}= 0.26$, $\Omega_{\Lambda} = 0.74$, $\Omega_{{\rm B}} = 0.04
63$ and $H_0=72\,{\rm km\,s^{-1}Mpc^{-1}}$.  The $\Lambda$CDM transfer
functions have been computed with 
{\sc cmbfast}~\cite{Seljak:1996is}. 
The transfer function for the WDM model has been adopted from \cite{bode}. 
The simulations were run with a box
size of  30 comoving $h^{-1}$ Mpc with $2\times 200^3$ gas and dark 
matter particles.

In Figure \ref{fig2} we show the fractional difference between the flux power spectra of 
a $\Lambda$WDM model  and  that of a $\Lambda$CDM model for a range of WDM
particle masses at $z=2.5,3,4$. The free-streaming
of the WDM particles has little effect for  particles with masses 
$\gsim 1$ keV (thermal WDM) at scales for which the linear matter power spectrum can
be reliably inferred (less than  5\%  at z=2.5 increasing to 
15\% at z=4).  For smaller masses of the WDM particle there is,
however,  a dramatic reduction of the flux power spectrum.   
Note that at small  small scales $\lsim 0.1$ s/km, the flux power
spectrum  shows a bump.  This is the effect of the  non-linear
evolution of the matter distribution which results in a transfer  
of power from large scales to small scales \cite{nara}.  
The diamonds in  Figure \ref{fig2}  show the   effect of adding 
randomly oriented velocities drawn from a Fermi-Dirac
distribution in the initial conditions for the model 
with $\mwdm$=1 keV. We confirm the result of \cite{nara} 
that the  effect of adding these free-streaming velocities 
on the flux power spectrum  is small,  of the order of 2 \% for
$k<0.03$ s/km. We will neglect the effect in the following.

\begin{figure}[]
\includegraphics[angle=0,width=8.0cm,height=8.0cm]{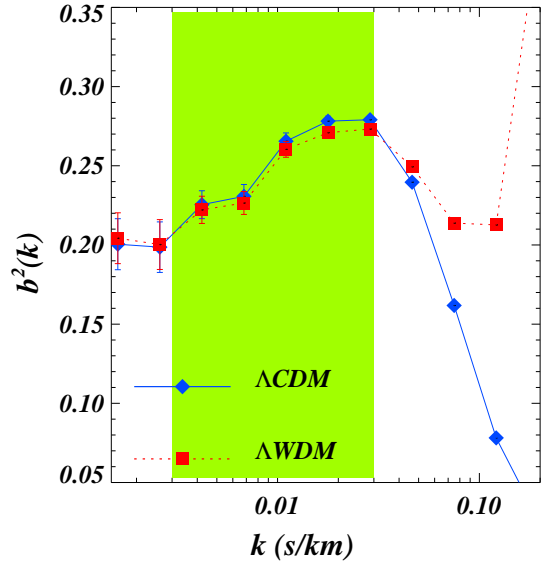}
\caption{\label{figbias} 
Bias function $b^2(k)$ between the flux power spectrum and the linear
dark matter power spectrum, $P_F(k) = b^2(k)\, P(k)$, for two cosmological
models: $\Lambda$CDM (diamonds) and $\Lambda$WDM (squares), with a
(thermal WDM) particle mass
$m_{\rm WDM} = 1$ keV. The simulation has a boxsize of  60 comoving $h^{-1}$Mpc
and  $2\times400^3$ (gas and dark matter) particles (see Section
\ref{hydro} for further details on  other simulation
parameters). The shaded area indicates the range of wavenumbers used in
our analysis.}  
\end{figure}

Figures \ref{fig1} and \ref{fig2} suggest that the \lya forest data 
will provide a tight lower limit to the mass of WDM particles. We will 
quantify this limit in the next section using the linear dark matter
power spectrum in the
range $(0.003-0.03)$~s/km as inferred by VHS.  We thereby checked how
strongly the bias function used to infer the linear matter power
spectrum depends on the presence of WDM. For this purpose we have run
a simulation with a 60 $h^{-1}$ comoving Mpc box with $2\times 400^3$
gas and wark dark matter particles for a (thermal WDM) particle mass
$\mwdm = 1$ keV. In the relevant wavenumber range, the difference
between the bias function extracted from this simulation and that of a
$\Lambda$CDM model (B2 model in VHS) is less than 3\%, as we can see
from Figure \ref{figbias}.
For smaller WDM particle masses the difference is expected to become
larger but as we discuss in Section \ref{purewdmres} this will have little 
impact on our final results. 

\subsection{Systematics Errors}
There is a number of systematic uncertainties and statistical errors
which affect the inferred power spectrum and an extensive discussion
can be found in \cite{croft,gnedham,vhs}. VHS estimated the 
uncertainty of the overall {\it rms} fluctuation amplitude of matter 
fluctuation to be 14.5 \% with a wide range of different factors 
contributing.  

In the following we present a brief summary. The effective optical
depth, $\tau_{\rm eff}=-\ln <F>$ which is essential for the
calibration procedure has to be determined separately from the
absorption spectra. As discussed in VHS, there is a considerable
spread in the measurement of the effective optical depth in the
literature.  Determinations from low-resolution low S/N spectra
give systematically higher values than high-resolution  high S/N
spectra. However, there is little doubt that the lower values from
high-resolution high S/N spectra are appropriate and the range
suggested in VHS leads to a 8\% uncertainty in the {\it rms}
fluctuation amplitude of the matter density field (Table 5 in VHS).
Other uncertainties are the slope and normalization of the
temperature-density relation of the absorbing gas which is usually
parametrised as $T=T_0\,(1+\delta_b)^{\gamma-1}$.  $T_0$ and $\gamma$
together contribute up to 5\% to the error of the inferred fluctuation
amplitude. VHS further estimated that uncertainties due to the C02
method (due to fitting the observed flux power spectrum with a bias
function which is extracted at a slightly different redshift than the
observations) contribute  about 5\%. They further 
assigned a 5 \% uncertainty to the somewhat uncertain effect of
galactic  winds and finally an   8\%  uncertainty due the 
numerical simulations (codes used by different groups give somewhat
different results).  Summed in quadrature, all
these errors led to the estimate of the overall uncertainty of 14.5\%
in the {\it rms} fluctuation amplitude of the matter density field.

For our analysis we use the inferred DM power spectrum in the range
$0.003\,{\rm s/km} < k < 0.03\,{\rm s/km}$ as given in Table 4 of VHS.
Note that we have reduced the values by 7\% to mimick a model with
$\gamma=1.3$, the middle of the plausible range for $\gamma$
\cite{temperature}. Unfortunately at smaller scales the systematic
errors become prohibitively large mainly due to the large contribution
of metal absorption lines to the flux power spectrum \cite{vhs} (their
Figure 3) and due to the much larger sensitivity of the flux power
spectrum to the thermal state of the gas at these scales.

\section{Bounds on the WDM mass from WMAP and the \lya forest}

In this section we will show the results of our combined analysis of
CMB data and \lya forest data.  The CMB power spectrum has been
measured by the WMAP team over a large range of multipoles ($l<800$)
to unprecedented precision on the full sky \cite{wmap}. For our
analysis we use the first year data release of the WMAP temperature and
temperature-polarization cross-correlation power
spectrum.

For the range of parameters $m_x$ and $\omega_x$ considered here, CMB
anisotropies are not sensitive to the velocities of WDM particles and
can therefore not discriminate between WDM and CDM models. It
is nevertheless crucial to combine the CMB and \lya data in order to get some
handle on other cosmological parameters. Since all recent CMB
experiments point toward a flat Universe with adiabatic scalar
primordial fluctuations, we adopt these results as a prior. We treat
the three ordinary neutrinos as massless (including small masses would
not change significantly our results). We use the publicly available
code {\sc camb}~\cite{Lewis:1999bs} in order to compute the
theoretical prediction for the $C_l$ coefficients of the CMB
temperature and polarization power spectra, as well as the matter
spectrum $P(k)$.  We derive the marginalized likelihood of each
cosmological parameter with   Monte Carlo Markov Chains as implemented
in the publicly available  code {\rm CosmoMC}~\cite{cosmomc}. The
calculations were performed on the Cambridge {\sc
cosmos} supercomputer. We extended the {\rm CosmoMC} package in order
to include the \lya forest data in a way similar to Ref.~\cite{vwh}.

\subsection{Lower bounds on $m_x$ in a pure $\Lambda$WDM Universe.}
\label{purewdmres}
We first compute numerically the $\Lambda$CDM matter power spectrum,
and then use equation~(\ref{Tshape}) with $\nu=1.2$ to obtain the pure
$\Lambda$WDM matter power spectrum leaving the break scale $\alpha$ as
a free parameter.  Apart from $\alpha$, our parameter set consists of
the six usual parameters of the minimal cosmological model: the
normalization and tilt of the primordial spectrum, the baryon density,
the dark matter density $\omega_\mathrm{DM}$ (equal here to
$\omega_x$), the ratio of the sound horizon to the angular diameter
distance, and the optical depth to reionization. Note that the Hubble
parameter and the cosmological constant can be derived from these
quantities, as well as $\sigma_8$ (the parameter describing the
amplitude of matter fluctuations around the scale
$R=8\,h^{-1}\mathrm{Mpc}$). In addition to these seven cosmological
parameters (with flat priors), we vary the parameter $A$ (see \cite{vhs,vwh})
which accounts for the uncertainty of the overall amplitude of the
matter power spectrum inferred from the \lya forest data (with a
Gaussian prior as described in \cite{vwh}).

\begin{table}[!ht]
\begin{tabular}{ccc}
\hline
     \hspace{0.6cm}  &  $\Lambda$WDM \hspace{0.6cm} &  $\Lambda$CWDM \\
\hline

$\Omega_x h^2$ & $0.124 \pm 0.015$ & $0.149 \pm 0.019$ \\

$\Omega_{\rm B}h^2$ & $0.024\pm 0.001$ & $0.024\pm 0.001$ \\

$h$ & $0.72\pm0.06$ & $0.71\pm0.06$ \\

$\tau$ & $0.18\pm 0.09$& $0.17 \pm 0.08$ \\

$\sigma_8$ & $0.96 \pm 0.08$ & $0.86 \pm 0.09$ \\

$n$ & $1.01\pm 0.04$ & $1.00 \pm 0.04$ \\

$\alpha~(\mathrm{Mpc}/h)$ & $0.06 \pm 0.03$ &  ----- \\

$f_x$ & ----- & $0.05\pm 0.04$ \\

\hline
\end{tabular}
\caption{The marginalized results for the recovered cosmological
parameters for our WMAP + \lya runs. The left column is for 
pure $\Lambda$WDM models
while the second column is for mixed $\Lambda$CWDM
models (with a fixed WDM temperature corresponding to $g_*(T_D)=100$). 
The values correspond to the peaks of the marginalized
probability distributions. Errors are the 68\% confidence limits. }
\end{table}

\begin{figure}[ht!]
\begin{center}
\hspace{-0.5cm}
\includegraphics[angle=0,width=9cm]{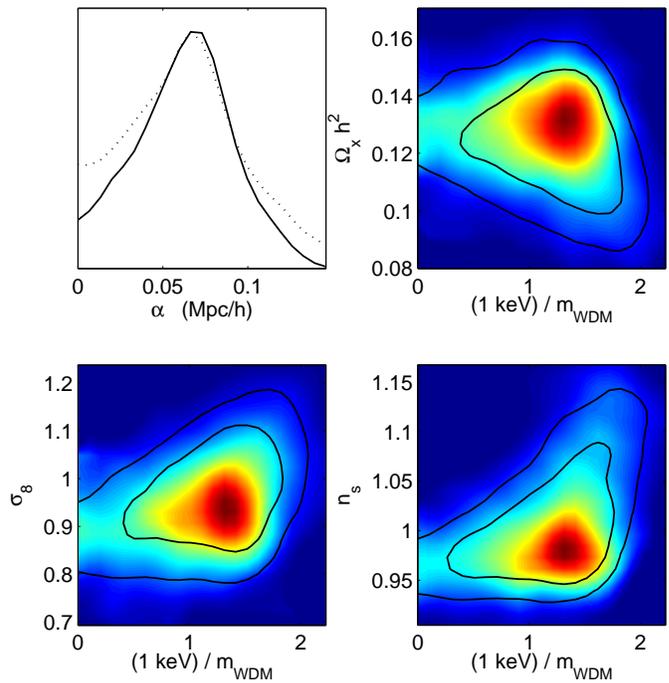}
\end{center}
\vspace{-0.5cm}
\caption{\label{fig3} Results for the pure $\Lambda$WDM model.
The top left panel shows the likelihood for the break scale 
$\alpha$. 
The other  three panels show the 1$\sigma$ and 2$\sigma$
contours for ($m_x^{-1}$, $\omega_x$), ($m_x^{-1}$, $\sigma_8$) and
($m_x^{-1}$, $n_s$), repectively, where $m_x$ is the mass of a {\it thermal} WDM particle. 
The solid curves correspond to the full marginalized likelihoods, while the
dashed curves and color/shade levels show
the mean likelihood of the samples in the Markov chains.  }

\end{figure}

The best-fit values for the cosmological parameters are 
summarized in the first column of Table 1.
In Figure \ref{fig3} (first panel), we show the marginalized
likelihood for the break scale $\alpha$. The probability peaks at
0.07~$h^{-1}$Mpc, but
the preference for a non-vanishing $\alpha$ is not statistically
significant, and our main result is the upper bound $\alpha
\lsim 0.11\, h^{-1} \mathrm{Mpc}$ at the 2$\sigma$ confidence level. Using
eq.~(\ref{fit}), we can derive the likelihood for the mass $m_x$, or
better, for its inverse, in order to keep a nearly flat prior. The
second panel of Figure \ref{fig3} shows the two--dimensional
likelihoods for the WDM parameters ($m_x^{-1}$, $\omega_x$), assuming
the WDM is a thermal relic. In this case the 2$\sigma$
bound on the mass is $m_x \gsim 550$~eV, while for a non-thermal sterile
neutrino as described in Section II this limit translates into $m_x \gsim
2.0$~keV.  

Note that for particle masses $\lsim$ 0.5 keV
the difference of the linear matter power spectrum of a WDM model
compared to that of the corresponding CDM model are not small anymore
at the relevant scales (Fig. 1). This makes the use of the linear bias
relation as described in Section III questionable when deriving the
likelihoods. The sensitivity to the mass of the WDM particle is
however so strong that this should have no effect on our conclusions
other than a possible small shift of our 1 and 2 $\sigma$ bounds in
this region.

We find no strong correlation between the mass of the WDM particle and
other cosmological parameters. This is expected as the characteristic
break due to the free--streaming length of the WDM particles cannot be
easily mimicked by the effect of other parameters.  Note however that
the constraints on $\Omega_x h^2$, $\sigma_8$ and the tilt $n_s$ of
the primordial spectrum are significantly weakened for WDM particle
masses $\lsim$ 1keV and that somewhat larger values of $\sigma_8$ and
$n_s$ become allowed.

\subsection{Upper bound on $m_x$ in a $\Lambda$CWDM Universe}

Let us now perform a similar analysis for the second model described
in Section II: a $\Lambda$CWDM Universe with a warm thermal relic
assumed to decouple when the number of degrees of freedom was $g_* =
100$ (as for instance for a light gravitino playing the role of the LSP).
We choose the same set of cosmological parameters as in the previous
subsection, except that the WDM sector is not described by the scale
$\alpha$ but by the density fraction $f_x = \omega_x /
\omega_\mathrm{DM}$ (where $\omega_\mathrm{DM} = \omega_x +
\omega_\mathrm{CDM}$).  The mass $m_x$ is proportional to the density
$\omega_x$ (Eq.~\ref{gravitino}), and the limit $m_x \longrightarrow
0$ corresponds to the standard $\Lambda$CDM model.  In this case,
we will therefore obtain an upper instead of a lower limit on the mass.

In Figure \ref{fig3} (first panel), we show the marginalized likelihood
for the WDM fraction. The 2$\sigma$ Bayesian confidence limit $f_x
\lsim 0.12$ can be translated into a limit on the mass, $m_x \lsim 16
\, \mathrm{eV}$. The other three panels show two--dimensional contours
for $m_x$ and the most correlated variables: $\sigma_8$, $\omega_X$ and
$n_s$.  There is now some correlation between the mass and the total
dark matter density, because increasing $\omega_\mathrm{DM}$ tilts the
small--scale matter power spectrum in such a way that it compensates to
some extent the red tilting introduced by $m_x$.  For the same reason,
one could expect some correlation with the tilt of the primordial
spectrum.  This does not occur because the value of the tilt is fixed
more precisely than that of $\omega_\mathrm{DM}$ by the CMB data.  The
gravitino model predicts globally less power on small scales than
$\Lambda$CDM. It therefore prefers smaller values of $\sigma_8$ in a
combined analysis with the CMB data on large scales. Precise
independent measurements of $\sigma_8$ could thus at least in principle
tighten the limit on the mass of a putative gravitino.

Note that in this analysis, we assumed $g_*(T_D)=100$, while as mentioned
in Section II
detailed studies of realistic supersymmetric models predict  $87 < g_*(T_D) <
104$, for the mass range investigated here ~\cite{Pierpaoli:1997im}. 
We have repeated our analysis assuming  $g_*(T_D)=90$
(which results in  a slightly smaller coefficient of proportionality
between the WDM mass and density) and found no significant difference.

\begin{figure}[ht!]
\begin{center}
\hspace{-0.5cm}
\includegraphics[angle=0,width=9cm]{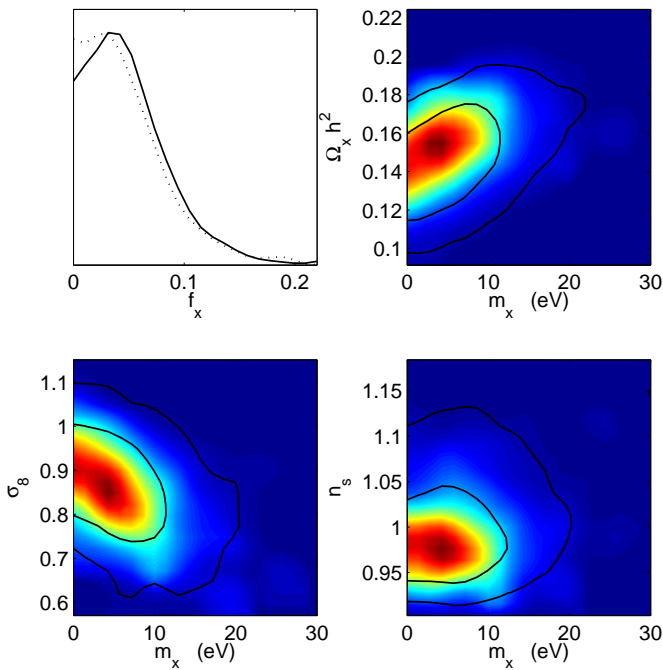}
\end{center}
\vspace{-0.5cm}
\caption{\label{fig4} Results for the $\Lambda$CWDM model with a light
gravitino (or a particle decoupling when $g_* = 100$).  The top left
panel shows the likelihood for the WDM density fraction $f_x$. Other
panels show the two--dimensional contours for the gravitino mass $m_x$
and the most correlated variables: $\sigma_8$, $n_s$ and $\omega_X$. The solid curves correspond to the
full marginalized likelihoods, while the dashed curves and color/shade
levels show the mean likelihood of the samples in the Markov chains.}
\end{figure}

\section{Conclusions}

We have obtained constraints on the mass of thermal and non-thermal 
WDM particles using a combined analysis of the cosmic microwave background
data from WMAP and the matter power spectrum inferred from a large sample 
of \lya forest data.

Assuming that the dark matter is entirely in the form of  warm
thermal relics, we find a 2$\sigma$ upper bound $m_x \gsim 550 \,
\mathrm{eV}$ on the mass of the DM particle.  This confirms the result of
Ref.~\cite{nara} who got a slightly bigger lower limit.  We have used
a significantly larger sample of QSO spectra, applied an improved
analysis which used state-of-the-art hydro-dynamical simulations and
taken into account a wide range of systematic uncertainties giving us
confidence that our result is robust.  Pushing this lower limit
to larger masses will require accurate modeling of the Lyman-$\alpha$
forest data at smaller scales. This is difficult due to the
sensitivity of the flux power spectrum to metal absorption and the
thermal history of the gas at these scales.

For non-thermal WDM particles like {\it e.g.} a sterile neutrino (with
the same phase-space distribution as that of a standard neutrino,
modulo a global suppression factor) this limit translates into $m \gsim
2.0 \, \mathrm{keV}$ (2$\sigma$). Note that in this case there exists
also an upper bound $m \lsim 5 \, \mathrm{keV}$, from limits on the
sterile neutrino decay rate in dark matter halos set by X-ray
observations \cite{Abazajian:2001vt}.  It has been noticed that
reionization at $z\sim 6$ requires $\mwdm \magcir 1$ keV, in addition if
Swift discovers $z \sim 10$ gamma ray bursts then this would require even
more stringent lower limits \cite{haiman}. The existence of a WDM
particle with a mass not much larger than these limits would increase
the inferred amplitude of the matter power spectrum $\sigma_{8}$ and
the inferred tilt of the primordial spectrum compared to a CDM
cosmology.

Our results can be easily converted into a bound on a
possible tiny interaction between CDM and photons, using the
analysis of Ref.~\cite{Boehm:2001hm} (who showed that such an interaction
would affect structure formation in the same way as WDM).

We have also obtained constraints on the mass of thermal relics which
decoupled at temperatures of the order of GeV or TeV and on their
contribution to the total matter density. In this case the number of
degrees of freedom in the Universe at the time of decoupling
$g_*(T_D)$ was of order of one
hundred and the particles cannot make up all
the dark matter in the Universe -- otherwise they would have a mass of
order 100~eV which would contradict the limit discussed above.  A
light gravitino is a prime example of such a particle and would only
be viable if it coexisted with ordinary cold dark matter (or with
another form of warm dark matter particle with significantly larger
mass). In this case, for particles with $g_*(T_D)$ in the range from
90 to 100, we find a 2$\sigma$ bound $m_x \lsim 16 \, \mathrm{eV}$. So,
if the gravitino is the LSP, as suggested by
gauge--mediated susy breaking scenarios, the susy breaking scale is
limited from above,
\begin{equation}
\Lambda_\mathrm{\sc susy}
\simeq \left(  \sqrt{3} m_{3/2} M_p \right)^{1/2}
\lsim 260 \, \mathrm{TeV}~.
\end{equation}
This conclusion is robust and relevant for supersymmetry
searching at future accelerator machines as LHC.
Indeed, if we assume that after
decoupling the gravitino background was enhanced by the decay of the
NSP (Next--to--lightest Supersymmetric Particle), as it happens
in some scenarios \cite{Borgani:1996ag}, the total density of gravitinos
as well as
their velocity dispersion can only be enhanced, making the bound even
stronger. The only  possible way to evade this bound on
the susy breaking scale (still assuming that the gravitino is the LSP)
would be to assume some entropy production after gravitino decoupling,
as suggested in Refs.~\cite{Baltz:2001rq,Fujii:2002fv}.

\section*{Acknowledgements.} 
We thank G.F. Giudice for getting us interested in setting a bound on
the scale of supersymmetry breaking, as well as George Efstathiou,
Steen Hansen and Sergio Pastor for useful comments. This work is
partly supported by the European Community Research and Training
Network ``The Physics of the Intergalactic Medium''. The simulations
were done at the UK National Cosmology Supercomputer Center funded by
PPARC, HEFCE and Silicon Graphics / Cray Research. MV thanks PPARC for
financial support. JL thanks the University of Padova and INFN for
supporting a six-month visit during which this work was carried out.
MV thanks the Kavli Institute for Theoretical Physics in Santa Barbara,
where part of this work was done, for hospitality during the workshop
on ``Galaxy-Intergalactic Medium Interactions''.

\end{document}